\documentclass{article}
\usepackage{spconf,amsmath,graphicx,float}
\DeclareMathOperator*{\argmax}{argmax}

\title{AUDIO-VISUAL DECISION FUSION FOR WFST-BASED AND SEQ2SEQ MODELS}
\name{
\begin{tabular}{c}
Rohith Aralikatti*, Sharad Roy*, Abhinav Thanda*, Dilip Kumar Margam, \\Pujitha Appan Kandala, Tanay Sharma and Shankar M Venkatesan
\end{tabular}\thanks{* These authors have contributed equally.}
\thanks{Submitted for review to ICASSP 2020 on October 21st, 2019.}}
\address{Samsung R\&D Institute India, Bangalore}
\begin{document}
\ninept
\maketitle
\begin{abstract}
Under noisy conditions, speech recognition systems suffer from high Word Error
Rates (WER). In such cases, information from the visual modality comprising
the speaker's lip movements can help improve the performance. In this work, we
propose novel methods to fuse information from audio and visual modalities at
inference time. This enables us to train the acoustic and visual models
independently. First, we train separate RNN-HMM based acoustic and visual
models. A common WFST generated by taking a special union of the HMM
components is used for decoding using a modified Viterbi algorithm. Second, we
train separate seq2seq acoustic and visual models. The decoding step is
performed simultaneously for both modalities using shallow fusion while
maintaining a common hypothesis beam. We also present results for a novel seq2seq
fusion without the weighing parameter. We present results at
varying SNR and show that our methods give significant improvements over
acoustic-only WER.

\end{abstract}
\section{Introduction}
\label{sec:intro}

Audio-visual Automatic Speech Recognition (AVASR) is a way to improve the performance of ASR systems in noisy conditions since lip movements generated during speech are independent of noise in the acoustic signal.

[1] and [2] propose lipreading using Discrete Cosine Transform (DCT) features for Hidden Markov Model (HMM) based models similar to traditional HMM acoustic speech recognition systems. Several methods for audio-visual fusion for HMM based models[3] have been proposed - coupled HMMs [4][5], twin HMMs[6] and synchronous HMMs[7].

Visual speech recognition has seen a resurgence with the rise of deep learning, in the form of end-to-end models such as CTC based models and seq2seq models. Chung et al. [8] proposed the use of 3D convolution for word level lip reading. Assael et al. [9]  proposed the LipNet architecture, where they train a character based CTC model for sentence level lip reading on the GRID audio-visual corpus [10]. In [11], the authors improve upon the results of LipNet by using cascaded attention-CTC and a highway network layer. In [12], the authors propose V2P - a CTC model trained on phoneme sequences along with a phoneme WFST for decoding. They demonstrate 40.9\% WER on the large vocabulary LSVSR dataset. In [13] and [14], the authors propose end-to-end audio-visual speech recognition systems, where the models are trained on both audio and visual features simultaneously. In the first paper above, the authors use a 3D convolution-ResNet-BLSTM architecture using the CTC loss, while the second paper proposes a transformer based model trained using both CTC and seq2seq losses.

All of the audio-visual methods described above require tight coupling of audio and visual modalities i.e. they involve feature level fusion or mid-level fusion. This requires that the model must be trained using both audio and visual features as input. Traditional speech recognition models are trained on large amounts of audio data and collecting such large datasets for audio-visual case is time consuming and costly.

In this paper, we introduce two novel methods for audio-visual speech recognition which satisfy the following conditions:\\
1)	It should be possible to train acoustic and visual models independently. This allows us more flexibility, as we can use different training procedures for each modality. Since the quantity of audio-visual data is generally much smaller than audio-only data, this also allows us to train each modality on different datasets.\\
2)	It should be possible to integrate audio-visual fusion to existing audio-only ASR systems, without degrading their performance in noise-free conditions.\\
We propose decision level audio-visual fusion for the following
speech recognition systems - the first is the RNN-HMM hybrid model with a WFST
graph (Section 3) and the second is the seq2seq model with attention (Sections 4 and 5). These methods do not involve rescoring after decoding, instead fusion is done during the decoding step.

\section{DATASETS}
\label{sec:format}

We report results on two dataset in Table 1, the GRID audio-visual corpus[10] and our private dataset, which we will henceforth refer to as Bixby dataset. The Bixby dataset was collected from 773 American speakers (of four ethnicities - White, African American, Asian American and Hispanic) and consists of commands spoken by users to Bixby - Samsung's mobile voice assistant. Some examples are "Bixby, will it rain today?" and "Schedule a meeting for next Saturday." etc. The test set for the Bixby dataset consists of 12 speakers, and we have ensured that we have at least one male and one female speaker from each ethnicity.
Test speakers used for the GRID dataset consists of two males and two females.
\begin{table}[ht]
\centering
\caption{Datasets used in the paper}
\begin{tabular}{ccccc}
\hline
\textbf{Dataset} & \textbf{Train} & \textbf{Test} & \textbf{Utterances}  & \textbf{Vocab} \\
\textbf{Name} & \textbf{Speakers} & \textbf{Speakers} & \textbf{per speaker} & \textbf{Size} \\
\hline
Bixby & 773 & 12 & 480 & 29469 \\
GRID & 30 & 4 & 1000 & 51 \\
\hline
\end{tabular}
\end{table}

\section{WFST FUSION FOR RNN-HMM MODELS}
\label{sec:pagestyle}

We train independent RNN-HMM models for speech recognition using both audio and visual inputs. These models are trained separately using the standard training procedures available in the Kaldi ASR toolkit [15]. For the RNN, we use a bi-directional LSTM with two layers trained using the nnet1 recipe in Kaldi.

\subsection{Feature extraction}
\label{ssec:subhead}

For audio, we use standard 13 dimensional MFCC features with delta and delta-delta extracted from 16 kHz wave files. A lip box is detected for every frame of the video using a YOLO [16] based detector. This data is augmented by flipping horizontally and adding artificial jitter to generate 8 videos for each original video. We use DCT features of the lip box with delta and delta-delta. Since audio has framerate of around 100 fps and video has around 25 fps, we up-sample the video features using 4x frame duplication.

\subsection{RNN-HMM models}
\label{ssec:subhead}

The RNN-HMM models for consist of two components - the neural network and the weighted finite state transducer (WFST) for decoding. The WFST is a composition of 4 separate FSTs - Language model (G), Lexicon (L), Phoneme Context (C), and HMM (H). The full WFST is generated by composition - HCLG = HoCoLoG. The RNN (bi-directional LSTM) is trained to output a probability vector over the states in the HMM for each frame. This vector is used to run standard Viterbi decoding on the complete WFST (HCLG). The best path from the generated lattice gives a sequence of words as output. The CLG cascade depends on the output vocabulary and can be changed depending on our target domain during testing. Both audio and visual RNN-HMM models can be pre-trained (possibly on different datasets).

\subsection{WFST fusion}
\label{ssec:subhead}
\begin{figure}[t]
\begin{minipage}[b]{.48\linewidth}
  \centering
  \centerline{\includegraphics[width=4cm]{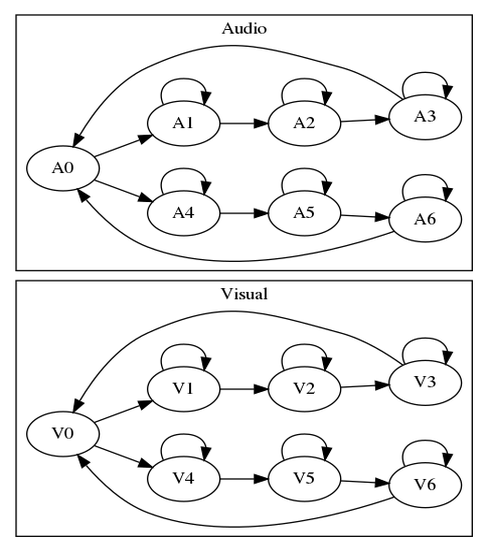}}
\end{minipage}
\begin{minipage}[b]{.48\linewidth}
  \centering
  \centerline{\includegraphics[width=4cm]{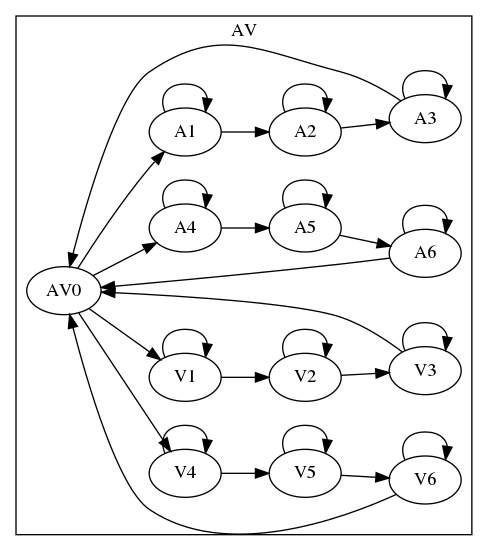}}
\end{minipage}
\caption{WFST fusion example. The audio and visual HMMs have two phonemes each (for representation). In the AV HMM each phoneme is repeated twice - once for audio and once for visual.}
\label{fig:res}
\end{figure}
The structure of the HMM component is such that each context dependent phone is represented by three states with self-loops (Fig. 1a). There is a common initial state (state 0) from which there is a transition to the first state of each context dependent phone. Along with self-loops, the first state has a transition to the second state and the second state to the third state. The third state has a transition back to state 0 thus making a loop of three states for each context dependent phone. We perform a special union of the audio and video HMMs followed by Kleene closure wherein we merge the state 0 of both HMMs. This makes the fused HMM have the same format as the individual HMMs and it has the three state loops from both HMMs (Fig. 1b).
We update the state and pdf identifiers of the video model such that they do not overlap with the audio states. We also make sure that the phoneme set being used by both the HMMs is the same so that we can safely compose this fused HMM with the common CLG component built on the output vocabulary.

\subsection{Modified Viterbi decoding}
\label{ssec:subhead}

The outputs of the audio and visual RNNs are concatenated and passed as input to a modified Viterbi decoding algorithm with the fused HCLG as described below. We use the same notations as given in [3]. $O=\{O_1,O_2,..O_T\}$ is the observed sequence (RNN outputs) for $T$ time frames. $\pi_i$ is the probability that the initial state is $i$, $b_i(O_t)$ is the probability of state $i$ generating $O_t$ and $a_{ij}$ is the transition probability from state $i$ to state $j$. $\delta_t(i)$ is the best score among all the beams which have state $i$ at time $t$, and $\psi_t(i)$ is the previous state in the beam given that the state at current time $t$ is $i$. We use $\psi$ to backtrack  and get the best hidden state sequence $Q^*=\{q_1^*,q_2^*,..q_T^*\}$ given the observed sequence. $P^*$ gives us the score for the best beam. Audio and visual HMMs  have N and M total states respectively.
\begin{figure}[t]
  \centering
  \centerline{\includegraphics[width=0.92\linewidth]{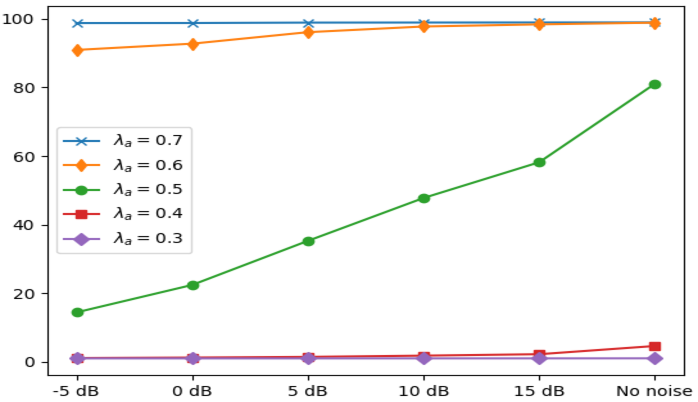}}
\caption{Percentage of audio HMM transitions vs noise (GRID). There is a large variation at $\lambda_a=0.5$ showing the self-weighing property.}
\label{fig:res}
\end{figure}
\begin{gather}
\lambda_i= \begin{cases}
		\lambda_a\quad\quad\quad 1\leq i\leq N \\
		1-\lambda_a\quad N<i\leq N+M \\
		\end{cases}
\end{gather}
\textit{1.\ Initialization}
\begin{gather}
\delta_1(i) = \lambda_i\pi_ib_i(O_1)\quad(1\leq i\leq N+M)\\
\psi_1 = 0
\end{gather}
\textit{2.\ Induction}
\begin{gather}
\delta_t(j)=\max\limits_{1\leq i\leq N+M}[\delta_{t-1}(i)a_{ij}]\lambda_jb_j(O_t)\\
\psi_t(j)=\argmax\limits_{1\leq i\leq N+M}[\delta_{t-1}(i)a_{ij}]\ \\(2\leq t\leq T,1\leq j\leq N+M)\notag
\end{gather}
\textit{3.\ Backtracking}
\begin{gather}
P^*=\max\limits_{1\leq i\leq N+M}[\delta_T(i)]\\
q_T^*=\argmax\limits_{1\leq i\leq N+M}[\delta_T(i)]\\
q_t^*=\psi_{t+1}(q_{t+1}^*)\quad\quad\quad\quad (1\leq t\leq T-1)
\end{gather}
While concatenating, we can individually give weights ($\lambda_i$) to the audio and visual RNN outputs, however it is observed that the optimal weight lies around $\lambda_a=0.5$ under all noise conditions. This means that our method is self-weighting, in other words, under high SNR audio transitions will have higher probabilities and under low SNR video transitions will have higher probabilities. This behaviour is observed by counting the total number of audio and video states in the beam during decoding. We see that $\lambda_a=0.5$ gives the highest variation in percentage of audio states as SNR increases (Fig. 2).

\section{TRAINING DETAILS FOR SEQ2SEQ MODELS}
\label{sec:typestyle}

The common training procedure specified below is used to train both audio and visual models. We use ADAM optimizer [17] with a maximum learning rate of 0.001. The first five epochs are the pre-training phase. Here we employ gradual learning rate warmup (learning rate is slowly increased from 0.00001 to 0.001, as done in [18] and [19]) and we train only on sentences whose length does not exceed 75 characters. We also utilize scheduled sampling [20], and we gradually increase the sampling probability from epochs 6 to 15. The sampling rate starts with a value of zero (epoch 6) and reaches a maximum value of 0.3 (epoch 15) and stays constant hereafter. We also employ learning rate scheduling after the pre-training phase. We divide each epoch into 5 sub-epochs. We scale the learning rate by 0.9 whenever sub-epoch validation cost has a relative improvement which is smaller than 1 percent. Early stopping with a delay of 5 epochs is used.

\subsection{Data augmentation}
\label{ssec:subhead}

We use the following data augmentation techniques during training. For the audio modality, we add noise to the training samples with a probability of 0.5. The noise is added to the signal so that resulting SNR (chosen randomly from a uniform distribution) lies between 0 dB and 15 dB. We use noise files from the DEMAND dataset [21] for augmentation. Apart from increasing the quantity of training data, it also ensures that the audio model has some robustness to noise. This is important, as we would like to show the improvement offered by fusion of the visual modality on an audio model that has seen noisy input during training. For the visual modality, we jitter the detected lip box by a small random pixel shift (between +5 and -5 pixels) both horizontally and vertically. The same random pixel shift is applied to all lip frames in an utterance.

\subsection{Pre-training}
\label{ssec:subhead}

For the GRID dataset, weights are initialized from the model trained on the Bixby Commands dataset. This is because the GRID dataset has only 33,000 utterances in all. Fine-tuning the Bixby Commands model on the GRID dataset gave significantly better WER for both audio and visual modalities, as compared to training the model purely on GRID utterances.

\subsection{Input features}
\label{ssec:subhead}

The audio model is trained on 23-dimensional filterbank features, extracted using Kaldi [15]. The window size is 25 ms and shift of 10 ms is used. The input to the video model is the sequence of detected lip images, which are resized to 50x100 resolution. Global mean-variance normalization is used for the image pixels, while the audio features are subjected to utterance level mean normalization.

\subsection{Model architecture}
\label{ssec:subhead}

The audio model has an encoder with three bi-directional LSTM layers, with 256 forward and backward units in each layer. The decoder has two unidirectional LSTM layers, with 512 units each. We use a time-factor reduction of two per encoder layer, similar to LAS [22]. The visual model has a convolutional front end, with two 3D convolution layers followed by two 2D convolution layers.  The encoder and decoder for the visual model are similar to the audio model, except that we do not use time factor reduction. Since video is at 25 fps (as compared to 100 fps for audio features), the visual model performed well even without time factor reduction. We use Luong Attention [23] for both modalities, so that the decoder can attend to the appropriate section of the encoded representation while producing the output.

\subsection{Other implementation details}
\label{ssec:subhead}

The seq2seq models are implemented and trained in Tensorflow [24]. The seq2seq models are trained to produce a character level output. The number of output classes are 29, and they include the 26 alphabets, $<$space$>$, $<$start$>$ and $<$end$>$ tokens.

\section{AUDIO-VISUAL FUSION FOR SEQ2SEQ MODELS}
\label{sec:print}

The two methods used for audio-visual fusion are described below.

\subsection{Log probability interpolation (shallow fusion)}
\label{ssec:subhead}

The typical decoding step involves a beam search where the decoded output is the sequence of output classes that maximize the probability distribution computed by the network.
\begin{gather}
y^*=\argmax\ P(y|x)=\argmax\ log(P(y|x))
\end{gather}
We modify this equation so that the quantity to be maximized at every beam search step is a weighted sum of the log probabilities of the audio and visual models.
\begin{gather}
y^*=\argmax(\lambda*P_a(y|x_a)+(1-\lambda)*log(P_v(y|x_v)))
\end{gather}
$\lambda$ is the weight given to the audio model and can be varied from 0.0 to 1.0 to control the relative contribution of each modality. In the results section, we show that given the right value of $\lambda$, the modified decoding step above results in better WER at various SNR levels (from -5 dB to 15 dB). This is similar to shallow fusion of language models used in machine translation and speech recognition ([25] and [26]).
\subsection{Lambda-free fusion}
\label{ssec:subhead}
\begin{figure}[t]
  \centering
  \centerline{\includegraphics[width=0.92\linewidth]{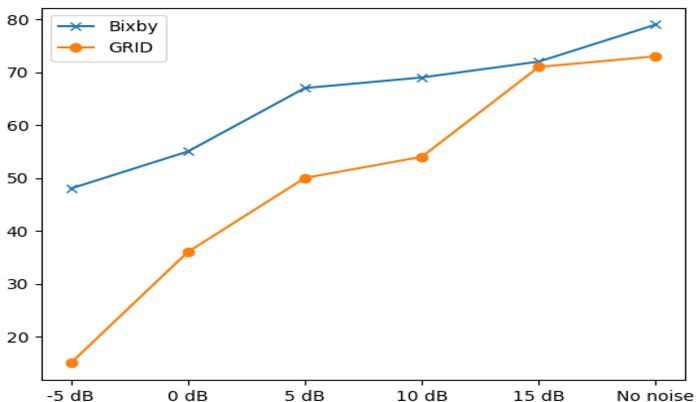}}
\caption{Percentage of candidates in the beam resulting from the audio model at different SNR values for $\lambda$-free fusion. As SNR reduces, the visual model contributes to more candidates and reduces WER. }
\label{fig:res}
\end{figure}
Fusion using the parameter $\lambda$ requires accurate estimation of SNR during inference (As seen from the results table, the value of $\lambda$ depends on the SNR). However, this may require a separate module/algorithm and is not ideal. Audio-visual models that are trained end-to-end on both modalities simultaneously do not suffer from this problem. Fusion using $\lambda$ involves ranking candidates after interpolating scores of both audio and visual models. Instead of ranking the output candidates in the beam based on either the audio or the visual model scores, we combine the scores from both modalities using an element wise max operation. By taking the element wise maximum of the scores prior to ranking, we ensure that classes that are given a high probability by either modality is present in the beam.

In Fig. 3, we plot the percentage of candidates in the beam that are present due to the audio model. We plot the average values of this percentage vs SNR. As SNR decreases, we see that the percentage reduces as expected and the visual model contributes more candidates to the beam, resulting in better WER.

\section{RESULTS}
\label{sec:page}

We test the performance of the two fusion methods at five different SNR values (-5 dB to 15 dB at intervals of 5 dB each).

\subsection{WFST fusion}
\label{ssec:subhead}

For decoding, we use the modified version of the Viterbi decoding implementation with beam width of 80 and lattice size of 20. We use a simple 5-gram language model built on the all the sentences in the dataset for scoring. We add artificial restaurant noise (PRESTO) from the DEMAND dataset [21] to the test set at varying SNR values. Since it is observed that the optimal value of $\lambda$ is around 0.5 for all noise conditions, we report all results with $\lambda=0.5$. From tables 2 and 3, we see the average relative improvement over the audio only WER is 18.76\% for the Bixby dataset and 47.10\% for the GRID dataset.
\begin{table}[ht]
\centering
\caption{Results (WER) for WFST fusion on Bixby Dataset}
\begin{tabular}{cccc}
\hline
\textbf{SNR} & \textbf{Audio} & \textbf{Visual} & \textbf{AV Fusion} \\
\hline
No Noise & 1.85 & 27.49 & 1.82 \\
15 dB & 2.58 & 27.49 & 2.75 \\
10 dB & 3.19 & 27.49 & 3.70 \\
5 dB & 7.73 & 27.49 & 7.09 \\
0 dB & 37.91 & 27.49 & 17.93 \\
-5 dB & 89.95 & 27.49 & 24.67 \\
\hline
\end{tabular}
\end{table}
\begin{table}[ht]
\centering
\caption{Results (WER) for WFST fusion on GRID Dataset}
\begin{tabular}{cccc}
\hline
\textbf{SNR} & \textbf{Audio} & \textbf{Visual} & \textbf{AV Fusion} \\
\hline
No Noise & 1.52 & 15.17 & 1.49 \\
15 dB & 9.38 & 15.17 & 6.33 \\
10 dB & 15.36 & 15.17 & 8.41 \\
5 dB & 25.64 & 15.17 & 11.08 \\
0 dB & 46.15 & 15.17 & 14.01 \\
-5 dB & 64.46 & 15.17 & 15.16 \\
\hline
\end{tabular}
\end{table}
\subsection{Seq2seq model}
\label{ssec:subhead}

A beam width of 20 is used during decoding. The noise used for testing (PRESTO noise in DEMAND dataset [21]) was not used for augmentation while training the audio model. For shallow fusion, at each SNR we estimate the best value of $\lambda$ using the validation set. 
From tables 4 and 5, we see that lambda-free fusion performs slightly worse as compared to shallow fusion. Considering all SNR values, shallow fusion provides an average relative improvement of 35.18\% (Bixby) and 32.42\% (GRID) as compared to audio only WER. Lambda-free fusion results in an average relative improvement of 25.17\% (Bixby) and 3.01\% (GRID).
\begin{table}[ht]
\centering
\caption{Results (WER) for seq2seq fusion on Bixby Dataset}
\begin{tabular}{ccccc}
\hline
\textbf{SNR} & \textbf{Audio} & \textbf{Visual} & \textbf{Log prob.} & \textbf{$\lambda$-free} \\
 &  &  & \textbf{interp.} & \textbf{Fusion} \\
\hline
No Noise & 8.92 & 27.91 & 7.59 ($\lambda=0.5$) & 8.43 \\
15 dB & 10.11 & 27.91 & 8.19 ($\lambda=0.5$) & 9.39\\
10 dB & 12.24 & 27.91 & 9.34 ($\lambda=0.5$) & 10.63\\
5 dB & 17.67 & 27.91 & 11.24 ($\lambda=0.5$) & 13.25 \\
0 dB & 35.78 & 27.91 & 17.25 ($\lambda=0.5$) & 20.42 \\
-5 dB & 74.65 & 27.91 & 25.86 ($\lambda=0.1$) & 31.84\\
\hline
\end{tabular}
\end{table}
\begin{table}[ht]
\centering
\caption{Results (WER) for seq2seq fusion on GRID Dataset}
\begin{tabular}{ccccc}
\hline
\textbf{SNR} & \textbf{Audio} & \textbf{Visual} & \textbf{Log prob.} & \textbf{$\lambda$-free} \\
 &  &  & \textbf{interp.} & \textbf{Fusion} \\
\hline
No Noise & 1.07 & 12.92 & 0.95 ($\lambda=0.9$) & 1.48 \\
15 dB & 1.5 & 12.92 & 1.38 ($\lambda=0.9$) & 2.18\\
10 dB & 2.31 & 12.92 & 1.97 ($\lambda=0.9$) & 2.97\\
5 dB & 5.22 & 12.92 & 3.75 ($\lambda=0.7$) & 4.89 \\
0 dB & 16.05 & 12.92 & 7.4 ($\lambda=0.7$) & 8.56 \\
-5 dB & 57.00 & 12.92 & 12.21 ($\lambda=0.5$) & 12.94\\
\hline
\end{tabular}
\end{table}
\section{CONCLUSION}
\label{sec:illust}

We have shown that audio-visual decision fusion methods work for seq2seq models and classical WFST-based systems. It should be possible to develop similar decision fusion methods for CTC based models. We can also consider trying to develop SNR estimation techniques based on fact that the percentage of audio candidates varies with SNR as seen in Fig. 2 and 3. Using such SNR estimation methods and an SNR-lambda lookup table (obtained from the validation set), it may be possible to use the ideal value of lambda for seq2seq fusion at runtime.

\section{REFERENCES}
\label{sec:refs}
[1] Yu, Keren, Xiaoyi Jiang, and Horst Bunke. ``Sentence lipreading using hidden Markov model with integrated grammar." International journal of pattern recognition and artificial intelligence 15.01 (2001): 161-176.\\\\\ 
[2] Puviarasan, N., and S. Palanivel. ``Lip reading of hearing impaired persons using HMM." Expert Systems with Applications 38.4 (2011): 4477-4481.\\\\\ 
[3] Rabiner, Lawrence R. ``A tutorial on hidden Markov models and selected applications in speech recognition." Proceedings of the IEEE 77.2 (1989): 257-286.\\\\\ 
[4] Nefian, Ara V., et al. ``A coupled HMM for audio-visual speech recognition." 2002 IEEE International Conference on Acoustics, Speech, and Signal Processing. Vol. 2. IEEE, 2002.\\\\\ 
[5] Pan, Hao, et al. ``A fused hidden Markov model with application to bimodal speech processing." IEEE Transactions on Signal Processing 52.3 (2004): 573-581.\\\\\ 
[6] Zeiler, Steffen, et al. ``Introducing the Turbo-Twin-HMM for Audio-Visual Speech Enhancement." INTERSPEECH. 2016.\\\\\ 
[7] Bengio, Samy. ``An asynchronous hidden markov model for audio-visual speech recognition." Advances in Neural Information Processing Systems. 2003.\\\\\ 
[8] Chung, Joon Son, and Andrew Zisserman. ``Lip reading in the wild." Asian Conference on Computer Vision. Springer, Cham, 2016.\\\\\ 
[9] Assael, Y. M., Shillingford, B., Whiteson, S., and de Freitas, N. ``Lipnet: End-to-end sentence-level lipreading." arXiv preprint arXiv:1611.01599 (2016).\\\\\ 
[10] Cooke, Martin, et al. ``An audio-visual corpus for speech perception and automatic speech recognition." The Journal of the Acoustical Society of America 120.5 (2006): 2421-2424.\\\\\ 
[11] Xu, Kai, et al. ``LCANet: End-to-end lipreading with cascaded attention-CTC." 2018 13th IEEE International Conference on Automatic Face \& Gesture Recognition (FG 2018). IEEE, 2018.\\\\\ 
[12] Shillingford, Brendan, et al. ``Large-scale visual speech recognition." arXiv preprint arXiv:1807.05162 (2018).\\\\\ 
[13] Petridis, Stavros, et al. ``End-to-end audiovisual speech recognition." 2018 IEEE International Conference on Acoustics, Speech and Signal Processing (ICASSP). IEEE, 2018.\\\\\ 
[14] Afouras, T et al. ``Deep audio-visual speech recognition” arXiv preprint arXiv:1809.02108.\\\\\ 
[15] Povey, Daniel, et al. ``The Kaldi speech recognition toolkit." IEEE 2011 workshop on automatic speech recognition and understanding. No. CONF. IEEE Signal Processing Society, 2011.\\\\\ 
[16] Redmon, Joseph, et al. ``You only look once: Unified, real-time object detection." Proceedings of the IEEE conference on computer vision and pattern recognition. 2016.\\\\\ 
[17] Kingma, D. P., and J. L. Ba. ``Adam: A method for stochastic optimization. arXiv 2014." arXiv preprint arXiv:1412.6980 (2014).\\\\\ 
[18] Chiu, Chung-Cheng, et al. "State-of-the-art speech recognition with sequence-to-sequence models." 2018 IEEE International Conference on Acoustics, Speech and Signal Processing (ICASSP). IEEE, 2018.\\\\\ 
[19] Zeyer, Albert, et al. "Improved training of end-to-end attention models for speech recognition." arXiv preprint arXiv:1805.03294 (2018).\\\\\ 
[20] Bengio, Samy, et al. "Scheduled sampling for sequence prediction with recurrent neural networks." Advances in Neural Information Processing Systems. 2015.\\\\\ 
[21] Thiemann, Joachim, Nobutaka Ito, and Emmanuel Vincent. "DEMAND: a collection of multi-channel recordings of acoustic noise in diverse environments." Proc. Meetings Acoust..2013.\\\\\ 
[22] Chan, William, et al. "Listen, attend and spell: A neural network for large vocabulary conversational speech recognition." 2016 IEEE International Conference on Acoustics, Speech and Signal Processing (ICASSP). IEEE, 2016.\\\\\ 
[23] Luong, Minh-Thang, Hieu Pham, and Christopher D. Manning. "Effective approaches to attention-based neural machine translation." arXiv preprint arXiv:1508.04025 (2015).\\\\\ 
[24] Abadi, Martín, et al. "Tensorflow: A system for large-scale machine learning." 12th {USENIX} Symposium on Operating Systems Design and Implementation ({OSDI} 16). 2016.\\\\\ 
[25] Gulcehre, Caglar, et al. "On using monolingual corpora in neural machine translation." arXiv preprint arXiv:1503.03535(2015).\\\\\ 
[26] Chorowski, Jan, and Navdeep Jaitly. "Towards better decoding and language model integration in sequence to sequence models." arXiv preprint arXiv:1612.02695 (2016).

\bibliographystyle{IEEEbib}
\bibliography{strings,refs}

\end{document}